\documentstyle[preprint,aps]{revtex}
\begin{document}
\title
{ 
  Study of a Nonlocal Density scheme for electronic--structure calculations
      }

\author{
Maurizia Palummo$^{(a)}$,
  Giovanni Onida$^{(a)}$,
Rodolfo Del Sole$^{(a)}$,
Massimiliano Corradini$^{(a)}$
}

\address{  $^{(a)}$ Istituto Nazionale per la Fisica della Materia -
 Dipartimento di Fisica dell' Universit\'a di Roma Tor Vergata\\
    Via della Ricerca Scientifica, I--00133 Roma,
   Italy}

\author{Lucia Reining$^{(b)}$}
\address{$^{(b)}$Laboratoire des Solides Irradi\'es,
   UMR7642 CNRS -- CEA,
   \'Ecole Polytechnique, F-91128 Palaiseau, France}

\par\noindent

\date{\today}

\maketitle
\begin{abstract}

An exchange-correlation energy 
functional beyond the local density approximation, based 
on the 
 exchange-correlation
kernel of the homogeneous electron gas and originally 
introduced 
by Kohn and Sham,
is considered for electronic structure calculations of 
semiconductors 
and atoms. 
Calculations are carried out for diamond, silicon, silicon carbide 
and 
gallium arsenide. The lattice constants
and gaps show a small 
improvement with respect 
to the LDA results.
 However, the corresponding corrections to 
the total energy of the isolated atoms
are not large enough to yield a substantial improvement for the 
 cohesive energy 
 of solids, which remains hence overestimated as in the LDA.

\end{abstract}
\newpage

\section{Introduction}  
Calculations based on the Kohn-Sham \cite{KS} 
formulation of density functional theory (DFT) \cite{HK} have 
become a  
prominent tool in condensed 
matter physics. Current work is dominated by local density 
approximation (LDA) studies, 
in which the exchange-correlation 
potential is a local function of the density \cite{KS,LDA}. 
\par 
Despite its enormous success, the local  density 
approximation has some shortcomings which motivated the 
increasing interest in approaches beyond it, like the Generalized 
Gradient Approximation
(GGA) \cite{GGA,Scheffler}, the average-density approximation 
(ADA), and 
the weighted density approximation (WDA) \cite{GJL}. 
However, at present
 none of these methods has replaced the simpler local density 
scheme, 
because they do not always yield systematic and  
consistent improvements with respect to the LDA, and because, 
except 
the GGA, 
they could not be implemented 
in a computationally tractable fashion. 
\par
It is therefore worthwhile to carry out electronic structure 
calculations 
using other approximations for the exchange--correlation energy 
functional. 
Kohn and Sham \cite{KS} proposed, together with the LDA, a 
correction to it which is exact up to the second order 
in the density fluctuations 
with respect to the average electron density $n$; 
hence it is appropriate for weakly inhomogeneous systems. 
It involves, as a basic ingredient, the exchange correlation kernel 
of the homogeneous electron gas, $K_{xc}^{heg}({\vec r}-{\vec r'})$, 
describing the change of the exchange-correlation potential at 
${\vec r}$ induced by a density change at ${\vec r'}$.
This functional was ruled out 
since Gunnarson and al. \cite{GJL} showed that 
different choices of how to include higher order terms may 
lead to very 
different results for the total energy in strongly inhomogeneous 
systems.
\par
In fact this
functional has never been applied to real solids,
while the LDA became the state-of-the-art of total-energy 
and band-structure 
calculations. 
Keeping into account that several realistic
parametrizations of $K_{xc}^{heg}$ based on Quantum Monte Carlo 
(QMC)
calculations are now available, we think that the relative    
 simplicity of this non--LDA (NLDA) functional, and its 
avoiding gradient expansions to treat inhomogeneity, 
suggest a further attempt to apply it to solids. 
\par
In Section II we describe in detail the NLDA functional and 
we propose a derivation which eliminates some arbitrariness 
in the treatment of higher--order density fluctuations. 
The choice which we adopt 
minimizes the error in the case 
of strongly inhomogeneous systems. In section III 
we describe the calculations performed for bulk silicon, 
diamond, silicon carbide and gallium arsenide, and the results of 
total 
energy 
calculations for  
atoms and pseudoatoms, as well as for a surface. 
The conclusions are drawn in Section 
IV.
\par
\vskip 2 truecm 
\section{NLDA exchange-correlation energy functional}  
\par
     In this Section we derive a correction
 to the LDA exchange--correlation
({XC}) energy for a weakly inhomogeneous system. Although 
the final 
result
will be coincident with a NLDA XC functional introduced by Kohn 
and Sham \cite{KS}, the
derivation given here is  different. It suggests a wider 
validity
range of the resulting functional, and renders the choice of how
to include higher orders less arbitrary.
\par
Let us consider the exchange-correlation potential 
$V_{xc}(\vec{r})$ 
in a weakly
inhomogeneous system. We can write:
\par\noindent
\begin{equation}
    V_{xc}([n ],\vec{r})= V_{xc}^{LDA}(n(\vec r)) + \int d\vec{r'} 
K_{xc}(\vec{r},\vec{r'}) [n(\vec{r'})-n(\vec{r})] 
+ .....
\end{equation}
where 
\begin{equation}
K_{xc}(\vec{r},\vec{r'})=\frac{ \delta V_{xc}([n ],\vec{r})} 
{\delta n(\vec{r'})}
\end{equation}
\par\noindent
is the functional derivative of the XC potential with respect to the 
density.
The first term on the right-hand side of (1) is obtained assuming, 
as in the LDA, that 
the
electron density $n(\vec{r'})$ is constant, equal to $n(\vec{r})$. 
The second term is an expansion 
of $V_{xc}([n ],\vec{r})$ 
to first-order
in $[n(\vec{r'})-n(\vec{r})]$, while higher-order terms are 
neglected. 
Calculating $V_{xc}([n ],\vec{r})$
up to the first-order in $[n(\vec{r'})-n(\vec{r})]$, 
$K_{xc}(\vec{r},\vec{r'})$ 
can be approximated at the
zero-order level, namely by using the homogeneous electron-gas 
(HEG)
form:
\noindent
\begin{equation}
      V_{xc}([n],\vec{r})= V_{xc}^{LDA}(n(\vec r)) + \int d\vec r' 
K_{xc}^{heg} (\vec{r}-\vec{r'};\tilde n(\vec{r},\vec{r'})) [n(\vec{r'})-
n(\vec{r})] + ......
\end{equation}
where $K_{xc}^{heg}(\vec{r}-\vec{r'};\tilde n(\vec{r},\vec r'))$ 
is the { XC} kernel (2) for the HEG (jellium) of
density $\tilde n(\vec{r},\vec{r'})$. 
Strictly speaking,  $\tilde n(\vec{r},\vec r')$ should be taken 
equal to 
$n(\vec{r})$. However, any choice in the range between $n(\vec{r})$ and 
$n(\vec{r'})$ will not alter the resulting XC
potential to first order in $n(\vec{r})-n(\vec{r'})$.
The choice of 
$\tilde n(\vec{r},\vec{r'})$ will be discussed below.
\par
       The next problem to solve is to find the { XC}-energy 
functional $E_{xc}[n]$
whose functional derivative ${ \delta E_{xc}}$/${\delta n(r)}$ 
yields the { XC}-potential (3). 
If $\tilde n(\vec{r},\vec{r'})$ is a symmetric function of $\vec 
r$
and $\vec {r'}$,
we can
easily show that the functional 
\noindent
\begin{equation}
E_{xc}[n]=E_{xc}^{LDA}[n]-\frac{1}{4} 
\int d\vec{r} \int d\vec{r'} 
K_{xc}^{heg}
\vec({r}-\vec{r'};\tilde n(\vec{r},\vec{r'})) [n(\vec{r'})-
n(\vec{r})]^2
\end{equation}
\noindent 
yields the following { XC}-potential:
\noindent 
$$
V_{xc}([n],\vec{r})= V_{xc}^{LDA}(n(\vec r)) + \int d\vec{r'} 
K_{xc}^{heg}(\vec{r}-\vec{r'};\tilde n(\vec{r},\vec{r'})) [n(\vec{r'})-
n(\vec{r})] 
$$
$$
 -(1/4) \int d\vec r'' \int d\vec{r'}  
 \frac{  dK_{xc}^{heg}(\vec{r''}-\vec{r'}; 
 \tilde n(\vec{r''},\vec{r'}))}{dn}  
 \frac {\delta \tilde n(\vec{r''},\vec{r'})}
 {\delta n(\vec r)} {[n(\vec{r'})-n(\vec{r''})]}^2 . ~~~~ (3a)
$$
\par\noindent
The { XC}-correlation potential (3a) is coincident with our 
starting 
point (3) up to terms of first order in the
density variation $n(\vec{r'})-n(\vec{r})$. 
Hence the { XC}-energy functional (4) is consistent
up to the second order in $[n(\vec{r'})-n(\vec{r})]$
with the { XC}-potential (3). 
\par
The exchange-correlation energy functional (4) becomes equal to 
the NLDA of Kohn and Sham if the density argument in $K_{xc}^{heg}$,  
$\tilde n(\vec{r},\vec r')$, is replaced by the average electron 
density $n$.     
Previous derivations of it were based on a 
partial
summation of the gradient expansion \cite {KS}, or on assuming 
weak density
fluctuations around the average density \cite{Sham}. The 
present
derivation is based on less restrictive assumptions: considering in 
fact
that the XC kernel is generally of short range in $|\vec{r}-
\vec{r'}|$ 
(according to Hubbard's
expression \cite{Hubb}, we expect that it decays after a few 
reciprocal Fermi
wavevectors), the density variation $[n(\vec{r'})- n(\vec{r})]$ 
must be small {\sl within this
range}, in order to allow the neglect of higher-order terms. We 
refer to
this situation as to weak inhomogeneity on the scale of the 
non--locality
range of $K_{xc}^{heg}(\vec{r}-\vec{r'};n)$, or shortly as to weak 
inhomogeneity.
\par
     If we consider an inhomogeneous system, the 
density
argument in the HEG XC kernel in (4) has some arbitrariness: in 
fact, 
since
the expression is valid up to the second order in density 
fluctuations,
the XC kernel has to be correct to zeroth order in it. This means 
that in principle any
density close to $n(\vec{r})$ and $n(\vec{r'})$ can be used 
therein. The 
simplest 
choices 
which
preserve the symmetry in $\vec{r}$ and $\vec{r'}$ are: 
(i) $n[(\vec{r}+\vec{r'})/2]$ and 
(ii) $[n(\vec{r})+n(\vec{r'})]/2$.
Choice (ii) is consistent with the ansatz (1), where the LDA term
is understood as the zero order contribution of an expansion 
around 
constant density, and the second term  
is a linear expansion of
the
XC potential at $\vec {r}$ in the change of the density at any point
$\vec
{r'}$, from the value assumed in the LDA term, $n(\vec {r})$, to
the
actual value $n(\vec {r'})$. In a common Taylor expansion, the 
derivative would
be
calculated at the initial point $\tilde n(\vec r,\vec {r'}) = n(\vec 
{r})$. 
This choice would however not
satisfy the symmetry requirement on $\tilde n(\vec r,\vec {r'})$. 
Moreover, one can be
easily
convinced that the error is smaller if the derivative is calculated
at the
central point of the interval, namely at $(n(\vec {r})+n(\vec
{r'}))/2$. If
the function is truly linear, the derivatives at these two points are
obviously equal and both points yield the exact result. Calculating
the
derivative at the central point of the interval, however, yields the
exact
result also in the case of a quadratic function.
Choice (i), on the other hand, has also been discussed 
in Ref.\cite{GJL}, where it has been demonstrated that it can  
lead to
very different results in the cases of strongly inhomogeneous 
systems,
and even to infinite values of the XC energy. Choice (i) is in fact 
ruled out by 
our derivation, since $n(1/2(\vec r + \vec {r'}))$ has no reason to 
lie
in the density range of $n(\vec r)$ and $n(\vec {r'})$.
Since the failure of (i)
has however been proposed as an argument against the use
of the functional, 
we analyze below the reasons of such different results and we 
show that 
choice
(ii) not only is consistent with our derivation, but also minimizes 
the 
error in real systems
with respect to choice (i).
\par
        The main source of error is that $K_{xc}^{heg}(\vec{r}-\vec{r'};n)$ 
becomes of long range
when $n$ is vanishingly small. This is understood on the ground 
of
self-interaction arguments \cite{GJL} and occurs even in the 
simplest model of
$K_{xc}^{heg}(\vec{r}-\vec{r'};n)$, namely Hubbard's model 
\cite{Hubb}. 
No relevant error occurs in (4)
when both $\vec{r}$ and $\vec{r'}$ are in regions of small 
electron density, since in this
case the factor  $[n(\vec{r'})- n(\vec{r})]^2$ vanishes. 
Serious errors can instead occur
when $\vec{r}$ (or $\vec{r'}$) is in a region 
of small density, while $\vec{r'}$ (or $\vec{r}$) is in a region
of relevant density. If we make choice (i), 
$n[(\vec{r}+\vec{r'})/2]$ 
may happen to be also small,
and the resulting kernel of long range. A very big wrong 
contribution 
may
be added to the XC energy in this case. If, on the other hand, we 
make
choice (ii), the density appearing in the kernel, 
$[n(\vec{r})+n(\vec{r'})]/2$ 
is non
vanishing and $K_{xc}^{heg}(\vec{r}-\vec{r'};n)$ is of short range 
in all relevant cases. We can still get a
wrong contribution to the { XC} energy,
but certainly
smaller in absolute value than that of choice (i).

We adopt hence
choice (ii) in the functional (4). The residual error can be checked 
a
posteriori by comparing the expectation values over the electron
wavefunctions of the XC potentials of eq.(3) and (3a) (with 
$(n(\vec{r})+n(\vec{r'}))/2$ in place of $\tilde n(\vec{r},\vec{r'})$): if the assumption of weak
inhomogeneity is valid, they must be very close to each other. We 
verified that this is not the case for atoms, but it is true for the 
analyzed 
semiconductors.  The deep reason for this is
that electrons in valence or lower conduction states, which are of
interest in most electronic-structure calculations, have very small
probabilities of being found in regions of small charge density, 
where
wrong contributions to the { XC} functional can be generated 
as described above. 
In any case 
we used form (3a), 
which is fully consistent with our ansatz (4) for the XC energy.
\par
    The XC kernel $K_{xc}^{heg}(q;n)$ of the HEG (Fourier transformed 
from ${\vec r} - {\vec r'}$ to ${\vec q}$) is strictly related to the 
so--called
static local-field factor $G(q;n)$ :  
          $ K_{xc}^{heg}(q;n) = -(4\pi e^2/q^2) G(q;n)$ . 
We have carried out numerical calculations using different forms
of $G(q;n)$: i) a Hubbard-like form, i.e. 
 $G(q;n)=q^2/2(q^2+q_{TF}^2(n))$, where 
$q_{TF}(n)$ is the 
Thomas Fermi wavevector at density n;   ii) two  
more realistic models for $G(q;n)$, the first based on the
known asymptotic behaviors of $G(q;n)$ for small and large q, 
and on approximate calculations in between 
 --the model of Ichimaru and Utsumi \cite{ICHI} (IU)--, 
and the second based on  a 
parametrization of QMC data by
of Moroni and Senatore \cite{MORONI}, recently proposed by some 
of us
(CPOD) \cite{CORRA}. 
\par
The last two models of $G(q;n)$ differ mainly in the 
asymptotic behavior for large $q$ (which in the IU model fails to
reproduce the exact result, i.e.  $G(q;n) \sim q^2$),
 and in 
the presence, in the 
IU model,
of a logarithmic singularity for $q=2q_f$ ($q_f$ is the Fermi wavevector), 
which does not appear in 
the CPOD    $G(q;n)$. (For a more detailed discussion of these differences,
see Ref. \cite{CORRA}.)
The resulting { XC} kernels in real space 
are quite similar, as shown in 
Fig. 1, where they are plotted for $r_{s}=2$. In the 
first case,   
the presence  of oscillations 
due to the logarithmic 
singularity leads to a more 
difficult convergence 
of the integrals, without affecting very much the final results. 
Hence, in the following we adopt the 
CPOD parametrization.

\section{Results}

\subsection{Bulk systems}
       Self-consistent electronic structure  calculations were carried 
out
using pseudopotentials and a plane wave basis. 
We generated the pseudopotentials on  a numerical grid,
using the method proposed by Troullier and Martins 
\cite{MT} for Silicon and Diamond, while for Ga and As we 
followed the 
scheme 
introduced by Hamann 
\cite{Hamann,Fuchs}.
All the pseudopotentials were used in the fully separable 
representation 
of
Kleinman and Bylander\cite{kleinman}, after having checked that
no ghost--state is generated.
In the case of GaAs non linear core corrections were included 
\cite{NLCC}. 
Plane-wave basis sets with a cutoff of 18  
Rydberg for Silicon,  20 Ryd for GaAs, and 40 Ryd for Diamond 
and 
Silicon 
Carbide 
are needed to achieve good convergence. 
Ten special points \cite{MPACK} were used in Brillouin Zone 
integrations.
\par
The NLDA XC
functional (4) and the XC potentials (3) and (3a) were calculated 
by
integration in real space (on a grid of $16\times 16\times 16$ 
points 
for Si, GaAs and Diamond, and of $24\times 24\times 24$ 
points for SiC) , with a cutoff 
$r_c$ in 
$|\vec{r}-\vec{r'}|$. 
Fig. 2 shows the convergence of the total energy of Si (we plot 
the correction to the 
LDA value), 
as a function of $r_c$,   
for the two more realistic forms, IU and CPOD, of $K_{xc}^{heg}$.   
The results 
are 
very similar, but the integrals calculated with 
the second model converge faster  
with respect to $r_c$. 
Clearly the computational effort for the integration increases 
linearly with the 
number of grid points 
used to calculate the integrals. However, two factors contribute
in limiting the numerical effort requested by the present NLDA method:
\par\noindent i) the convergence with $r_c$ 
is fast, showing the short range character of
$K_{xc}^{heg}$
both for silicon and 
for 
the other materials considered;
\par\noindent ii) since, as we verified,
self-consistency effects are very small, all the calculations
can be done starting from the LDA
wavefunctions and converge to the true NLDA ground--state in a very 
small number of iterations.
\par
We
have checked the validity of the assumption of weak 
inhomogeneity, by
calculating, in the case of Silicon and Diamond,
the expectation values of both the XC
potentials (3) and (3a): they differ by less than 0.01 eV, showing 
that
terms of second order in the density variations are negligible in 
the XC
potential, and therefore that Si and Diamond are weakly 
inhomogeneous 
systems in
the
sense discussed in Section II.
The same is true for cubic SiC,  since 
the energy changes between NLDA and LDA are of the same order 
as 
in Si. However  in the case of GaAs, where the effect of the large 
core charge density 
is partially taken into account 
through the non linear core corrections, 
the differences are as large as 0.3 eV, which suggests that the 
density 
argument
should be chosen more carefully.
In principle  the pseudopotentials used  must be generated 
carrying out atomic calculations within the same NLDA scheme
as that used in
the solid calculation. However, it  would be 
simpler to employ the ready-to-use LDA pseudopotentials, under 
the assumption that exchange-correlation effects on the 
pseudopotential are scheme independent. 
We carried out our calculations using both pseudopotentials, 
generated within the LDA and the NLDA schemes.
The results, discussed 
below, turn out to be very similar.

In Table \ref{tableground_1} we show the ground state properties 
obtained for the 
materials examined: it is evident that all
relevant quantities change very little with respect to the LDA.  
The
equilibrium lattice constants increase slightly,   
 while the bulk modulus decreases for Si, C and SiC 
and increases for GaAs.  
The small changes 
of the ground state properties obtained for the materials 
studied
 are somehow comforting, in view of the fact that gradient 
corrections
sometimes overcorrect LDA results 
\cite{GAR,CORSO}.   
In a recent paper by Fuchs at al. \cite{Scheffler}, a 
new set of GGA results for different
solid compounds are reported, and the role of core-valence 
exchange-correlation is investigated.
These data show that when LDA-generated pseudopotentials are 
used with GGA XC functionals and  nonlinear
core corrections are ignored, the aforementioned overcorrection of 
GGA results with respect to LDA does not occur 
(see Tables V and VI of Ref. \cite{Scheffler} for diamond
and silicon, respectively). 
The overcorrections appear instead if the pseudopotentials are 
consistently 
generated within the GGA scheme.
In conclusion,  
when we generate consistently the pseudopotentials 
in the new NLDA scheme, the NLDA ground state properties remain very 
close to the LDA  
results, while overcorrections of the lattice constants occur in the 
case of consistent GGA calculations. 
\par 
For all the materials considered the NLDA electronic structures 
(calculated at the same lattice parameter using NLDA 
pseudopotentials)
show slight differences with respect to the LDA band structure 
with a 
general very small increase of the gaps.
The values of the main gaps with both the LDA and NLDA schemes 
are 
shown in Tab.\ref{tablegaps_2} 
\par
The fourth column of tab. \ref{tableatoms_3} shows the 
NLDA correction to the ground state total energy (per atom) of the solid,
which is always decreased with respect to the LDA, both using 
LDA or 
NLDA pseudopotentials.
For Silicon this energy lowering agrees with QMC results obtained 
by 
Fahy et al. 
in ref. \cite{FAHY}, while for
diamond
QMC yields a total energy slightly less negative than the LDA one, 
at
variance with our results \cite{FAHY}.
But in view of the uncertainty 
still present in the total energy 
obtained from different QMC calculations (mostly due to the usage 
of a 
pseudopotential generated within 
the DFT-LDA and to the variational character of the 
wavefunctions, see 
for example the 
results
shown in refs \cite{FAHY},\cite{LI},\cite{KRALIK} ), we
believe that the discrepancies
 of the order of a few tenths of an eV per atom are
acceptable. 
\par
For the silicon case we also calculated the NLDA total energy  
using 
Hubbard's model of $K_{xc}^{heg}(\vec{r}-\vec{r'},n)$. 
It is clear from eq. (4) that, in this case, the XC NLDA functional is 
positively
definite, since the Hubbard's $K_{xc}^{heg}$  is always negative. 
In fact we found an increase of the total energy per atom of 
about 1.5 eV. 
On the other hand it is reasonable that, if the total energy 
increases, 
the gap between the filled and the empty states closes. 
This is indeed our result: the direct gap using Hubbard's 
$K_{xc}$ 
is 0.6 eV  smaller than 
the LDA value. 
These very different results show that it is very important 
to use a good description of $K_{xc}$.
\\ 
\\ 
\subsection{Atoms}
\par 
In order to obtain the cohesive energy of solids 
one also has to calculate the  atomic energies using the same XC 
functional. This is a stringent test, since the condition of weak 
inhomogeneity is hardly fulfilled in atoms. We start carrying out 
all-electron calculations for a number of atoms. 
In Tab. \ref{tableerrors_4} we compare the results obtained for 
the 
ground state 
energy of some neutral atoms
in the  LDA and NLDA schemes with the "experimental" 
values based on 
the compilation of Veillard and Clementi 
(\cite{VECLE} as reported in ref.\cite{CARROL}), and with recent 
GGA 
results \cite{PASQUA}. 
The NLDA correction increases with the size of the atoms, but, 
at variance with 
the GGA method,  
it is generally not enough to correct the LDA underestimation of 
the 
binding energy.  
\par
The reason of this failure is evident in Fig. 3, where the NLDA 
$V_{xc}$ potential of 
the Neon atom is shown together with the LDA and GGA ones.
The differences are clear: there is a region, at intermediate 
distances from 
the nucleus, where the NLDA XC potential becomes more negative 
than the LDA potential, but at large distance it remains 
substantially equal, with no correction to 
the wrong (exponential) decay of the LDA $V_{xc}(r)$.
The GGA has this shortcoming too,
but compensates for it in the region near the nucleus, where
$V_{xc}^{GGA}$ is more attractive than our and the LDA $V_{xc}$.  
Whether or
not $V_{xc}^{GGA}$ is closer to the true exchange-correlation
potential than ours in this region is not clear at present. It is
possible that the good agreement between GGA and "exact" total
energies of atoms is a consequence of the cancelation of two
errors, namely the lack of the Coulombic tail at long range, and a
possible
overactractive behavior at small distances, the
reciprocal cancelation of these two errors being optimized by the
suitable choice of parameters
made within the GGAs. 
\par
In order to 
compute
the cohesive energies, both atoms and solids must be treated on the same 
footing, 
i.e. using the same
pseudopotential. Hence, 
we generated 
the pseudopotentials
including the NLDA exchange correlation energy and applied the 
NLDA
functional to pseudoatoms,
in order to have a consistent approach in the atomic and solid 
calculations.
In Fig.4 we report the Silicon pseudopotentials, within the LDA 
and the NLDA
scheme:  
the curves obtained in the two schemes look quite similar, 
the main differences being confined to the core region $r < 1$ a.u. 
In this region, which is not really important for the computation 
of 
matrix elements, being  
 weighted by the $r^2 dr$ volume element, 
the
NLDA pseudopotential
are slightly less attractive than the LDA one.
Less evident, but 
more important are the differences for $r > 1$ a.u.,
where the NLDA curves are slightly deeper than the LDA ones.
\par 
The ground state total energies obtained in 
the LDA and NLDA
schemes for the pseudoatoms  
considered here are compared in Tab.\ref{tableatoms_3}.  
  In our NLDA calculations 
the atoms are always treated as spin unpolarized systems, assuming 
that 
any 
contribution of spin polarization is well described at the local spin 
density (LSD) level. 
(This conjecture is substantiated by Fuchs et al.'s calculations 
\cite{Scheffler}, where spin-polarization energies calculated 
within GGA and LSD  differ by about 0.1 eV.)
Little improvement with respect to LDA is obtained for the total 
energies, so that the calculated cohesive energies, 
quoted in Table \ref{tableground_1}, do not improve 
substantially. 
\par
\subsection{Surfaces}
\par\noindent
Although the inclusion of the NLDA functional introduced here 
do not show strong 
differences with 
respect to the LDA
for the two limit cases, bulk and atomic systems, 
for the sake of completeness we also report 
briefly 
the results obtained for a surface.
\par 
We calculated the ground state energy and the equilibrium 
structure of 
the 
2x1 reconstruction of Si(100) using both the standard LDA and 
the 
NLDA functional. 
Using NLDA a decrease of the total energy of 0.12 eV/atom, 
which is of the same order as the one obtained in bulk silicon,
is obtained. 
The equilibrium geometry does not show 
any appreciable difference with respect to that obtained using the 
LDA 
exchange and correlation potential. 
\section{Conclusions}
An old non-LDA exchange-correlation functional, originally 
derived by Kohn and Sham for weakly inhomogeneous systems, 
has been implemented for usage in electronic structure 
calculations. A new derivation, and a well defined treatment of 
higher-order fluctuations, have been devised, suggesting that this 
functional should be reliable for systems with slow, rather than 
weak, 
density variations (i.e. with density variations smooth on the 
scale 
of the inverse Fermi wave vector).
NLDA results for the lattice constant  of Si, diamond, cubic SiC and 
GaAs
are  slightly 
better than the LDA ones when compared with experiment. The 
overcorrection characteristic of the Generalized Gradient 
Approximations 
is avoided.
Negligible changes with respect
to LDA have been found also for a surface, Si(100)2x1.

On the other hand, the application of the XC functional to atoms 
(pseudoatoms) improves the total
energies only partially (slightly) 
with respect to the LDA values. Hence, no substantial improvement 
comes out for the cohesive energy of solids. The reason for this 
failure  is probably due to the fact that the Coulombic
 long--range tail of 
$V_{xc}$, which is important in the case of atoms, is still lacking.

In general, we obtain changes with respect to LDA which are 
smaller 
than those needed to match the experimental values. This
is a consequence of the conservative ansatz made about the 
density 
argument in eqs. (3) and (4), meant to avoid long-range 
components of 
$K_{xc}$ and possible divergences of the XC energy.  
Even though this goal has 
been obtained, the NLDA part of the XC potential is generally 
underestimated. This leads to important discrepancies with 
experiments 
in the case of atoms, whose binding energies are not fully 
corrected with 
respect to the LDA. However, in some cases, e.g. for bulk 
semiconductors, the 
present description is closer to reality than other approaches. 

\section{Acknowledgements}

 We acknowledge useful discussions with  R. Resta and G. Senatore
 in the early stage of this work. 
 Computer resources on an Origin 2000
 system were granted by INFM and CINECA 
 (Interuniversity Consortium of the
 Northeastern Italy for Automatic Computing)
 under the  account cmprmpi0.
 We are grateful to S. Goedecker for providing us an efficient
 code  for Fast Fourier Transforms \cite{FFT1,FFT2}.
 This work was supported in part by the European Community 
 programme ``Human Capital and Mobility'' (Contract
 No. ERB CHRX CT930337).

\begin{figure}
\caption{XC kernels of the homogeneous electron gas, 
$K_{xc}(R)$, according
to the model of Ichimaru and Utsumi (full line) and to the 
parametrization of QMC data  of ref. {\protect\cite{CORRA}} 
(dashed line) 
,  as a
function of $R={r*k_F}$.}
\label{xckernels_fig1}
\end{figure} 
\begin{figure}

\caption{Convergence of NLDA 
correction as function of $r_c$ in bulk silicon, using 
the model of Ichimaru and Utsumi (full line) and 
the
parametrization of QMC data  of ref. {\protect\cite{CORRA}} 
(dashed 
line). }
\label{convergence_fig2}
\end{figure}
\begin{figure}

\caption{The XC LDA potential (full line) 
compared with the potential obtained including self-consistently 
the NLDA correction  
(dashed line) and with the GGA potential (dotted line)
for the Ne atom. }
\label{vxclda_fig3}
\end{figure}
\begin{figure}

\caption{ Unscreened pseudopotentials within the NLDA 
(dashed lines) 
and LDA (full lines) for Silicon. }
\label{pseudonlda_fig4}
\end{figure}

\newpage
\begin{table}
\caption{Ground state properties of the bulk silicon, 
diamond and silicon carbide and Gallium Arsenide   
 as obtained in LDA and NLDA calculations. 
The first column indicates the XC scheme used to generate the 
pseudopotentials, the 
second one the scheme employed for the XC energy of the solid. }
\begin{tabular}{ c |  c  c  c  c  c }                
& $potential$ & $E_{xc}$ & $a_0 (au)$ &$B_0 (Mbar)$ & $E_b 
(eV)$  
\\
\hline
& LDA & LDA & 10.17 & .98 &  5.31  \\ 
& LDA & NLDA & 10.19 & .97 &  5.33  \\
Si & NLDA & NLDA & 10.19 & .95 &  5.28 \\ 
& Exp. & & 10.26 & .99 &  4.63  \\ 
\hline 
& & & & & \\  
& LDA & LDA & 6.73 & 4.51 &  8.63 \\
& LDA  & NLDA & 6.74 & 4.44  & 8.49  \\
C & NLDA & NLDA & 6.75 & 3.93  & 8.46 \\
& Exp. & & 6.74 & 4.42 &  7.37  \\ 
\hline 
& & & & &  \\
& LDA & LDA & 8.15 & 2.25  &  7.42\\
& LDA & NLDA & 8.15 & 2.14  &  7.47 \\
SiC & NLDA & NLDA & 8.16 & 2.00 & 7.35  \\ 
& Exp. & & 8.24 & 2.3 & 6.34 \\ 
\hline 
& & & & & \\
& LDA & LDA & 10.55 & 0.77 &  4.00 \\
GaAs & LDA  & NLDA & 10.63 & 0.89 & 3.95 \\
& NLDA & NLDA & 10.60 & 0.85 & 4.01  \\ 
& Exp. & & 10.68 & 0.75 & 3.26 \\ 
\end{tabular}
\label{tableground_1}
\end{table}

\begin{table}

\caption{Main direct gaps (eV) for the bulk silicon,
diamond and Gallium Arsenide
 as obtained in LDA and NLDA calculations at fixed lattice 
parameter
( $ a_0$ = 10.16, 6.74 and 10.6 respectively).}
\begin{tabular}{ c | c  c  c  }                 
& & LDA & NLDA  \\
\hline
&$\Gamma$  & 2.57  &  2.60 \\
Si &$X$  & 3.57 & 3.63  \\
&$L$  &  2.86 & 2.90  \\  
\hline
&$\Gamma$  & 5.62  &  5.70 \\
C &$X$  & 11.21 & 11.32   \\
&$L$  &  11.32 & 11.39  \\  
\hline
&$\Gamma$  & .39 & .38  \\
GaAs &$X$  & 3.99 & 4.01  \\
&$L$  & 2.05 &  2.06 \\  
\end{tabular}

\label{tablegaps_2}
\end{table}

\begin{table}
\caption{List of total energy and cohesive energy changes 
with respect to the LDA results, 
due to the use of NLDA scheme for the considered materials.   
(the LDA spin correction for total energy of pseudoatoms is 
included).  
The first column indicates the XC scheme used to generate the 
pseudopotential, the 
second one the scheme employed for the XC energy of the pseudo 
atom.}
\begin{tabular}{  c |  c  c  c  c  c  }                 
&  potential & Exc & $\delta E_{at}$ & $\delta E_{sol}$ & $\delta
E_{coh}$ \\ 
\hline
& & & & & \\
Si & LDA & NLDA & -0.09 &  -0.11 &  + 0.02  \\
& NLDA & NLDA & -0.31 & -0.28 &  -0.03 \\  
\hline
& & & & & \\
C & LDA & NLDA & -0.57 & -0.43 & -0.14  \\
& NLDA & NLDA & -0.70  & -0.54 & -0.16 \\  
\hline
& & & & & \\
SiC & LDA & NLDA & &  -0.38 & +0.05 \\
& NLDA & NLDA & &  -0.57 &  -0.07 \\  
\\
\hline
& & & & & \\
GaAs  & LDA & NLDA & -0.35 (As); -3.51 (Ga) & -1.88  & -0.05   \\
& NLDA & NLDA &  -0.36 (As);  -1.72 (Ga)&  -1.03 &  +0.01 \\ 
& -0.005   \\
\end{tabular}

\label{tableatoms_3}
\end{table}

\begin{table}
\caption{Ground state total energy (Hartree) of some atoms
from He
to Argon, obtained
 in the LDA and NLDA schemes, and compared with the
experimental values.
Each calculated total energy includes the spin polarization
corrections
obtained in the LDA
scheme. }
\begin{tabular}{ c  c  c  c  c  }                 
g.s. energy (a.u.) & & & & \\
  & LDA & NLDA  & GGA & EXP \\ 
\hline
& & & &  \\
$^{4}Be$  & -14.44   & -14.51 & -14.64 & -14.67 \\
$^{6}C$   & -37.46  & -37.62  & -37.78 & -37.84  \\
$^{8}O$  & -74.50  & -74.81 & -74.99 & -75.07 \\
$^{10}Ne$  & -128.18  & 128.77 & -128.94 & -128.94  \\
$^{11}Na$  & -161.38 & -162.11 & -162.25 & -162.25 \\
$^{14}Si$  & -288.09 & -289.23  & -289.33 & -289.34  \\
$^{18}Ar$ & -525.65 & -527.59 &  -527.54 & -527.54 \\
& & & &  \\
\end{tabular}
\label{tableerrors_4}
\end{table}

%
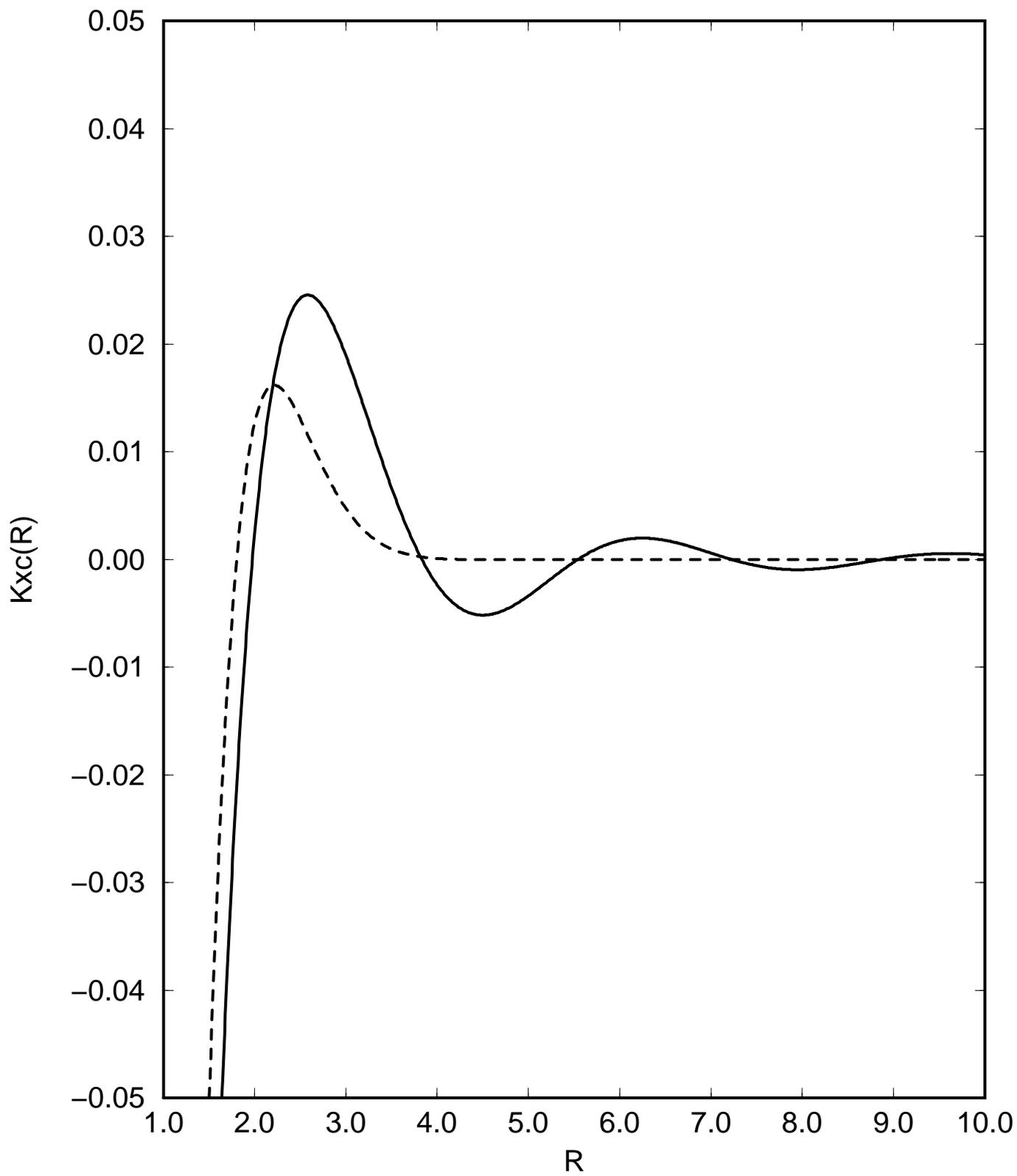
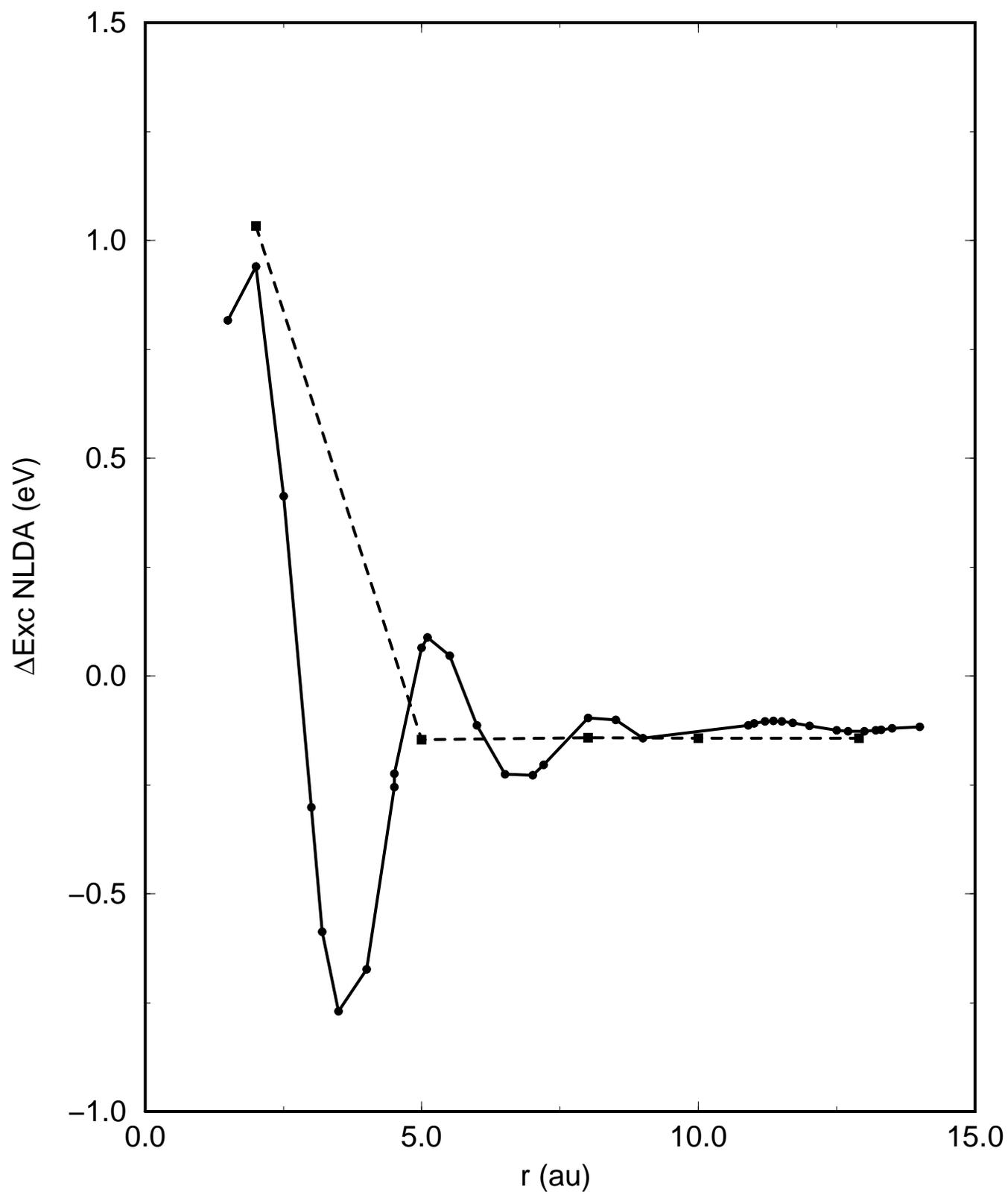
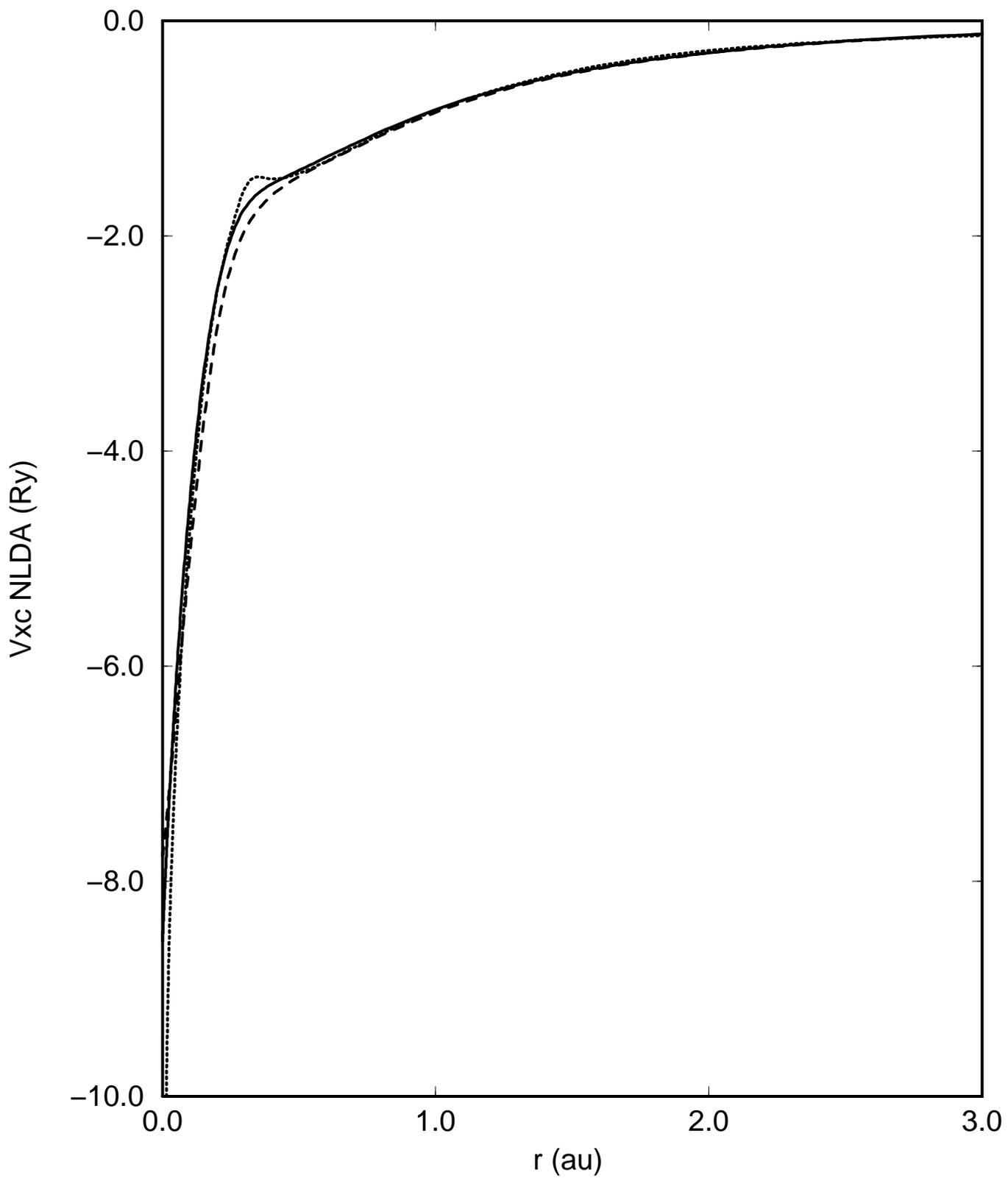
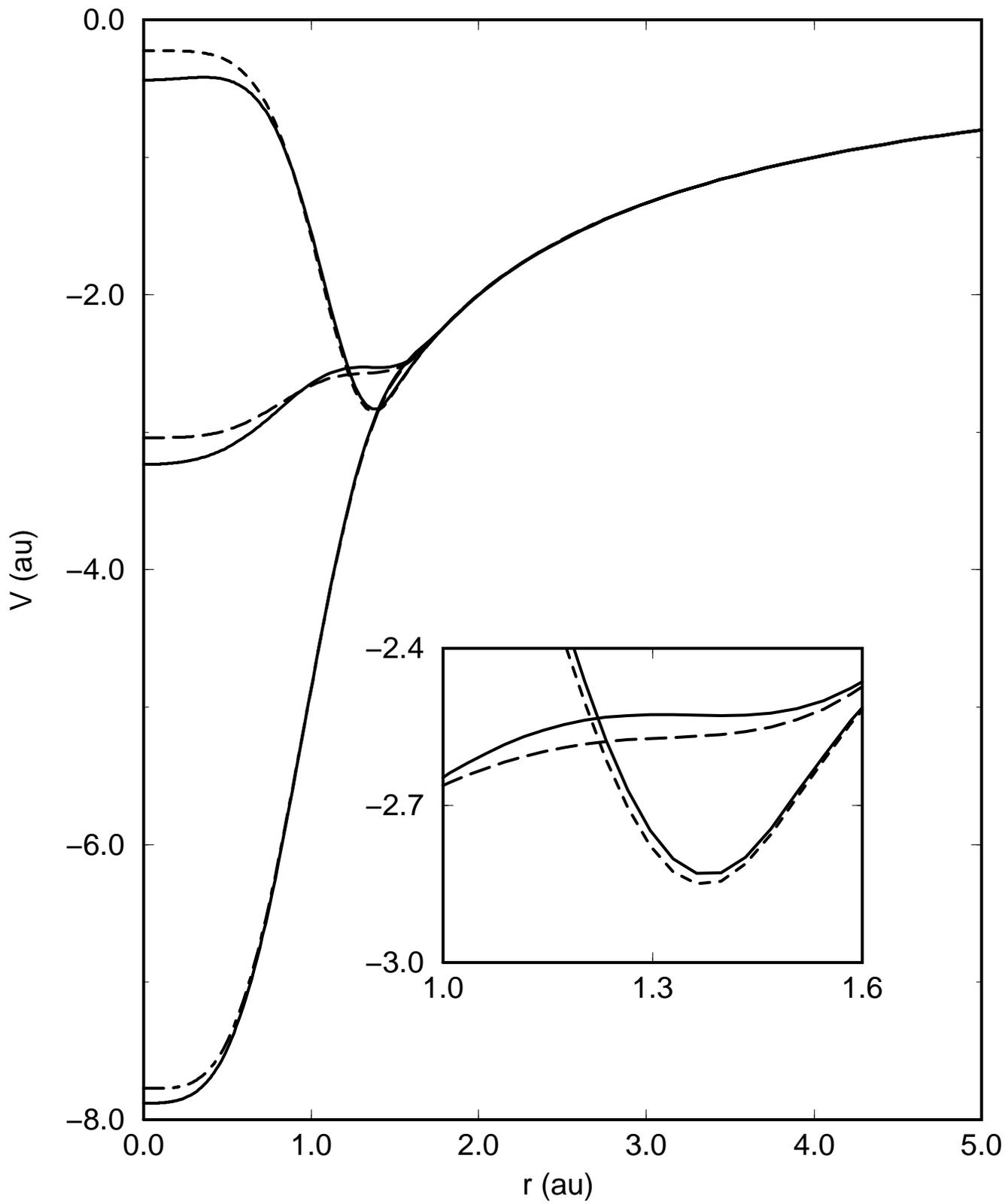
\end{document}